\documentclass[aps,prl,nofootinbib]{revtex4-1}

\usepackage{amssymb,graphics,graphicx}
\usepackage{footnote}
\usepackage{fullpage}

\usepackage{amsmath, amsfonts}

\begin{document}

\title
{\Large \bf USING NEWTON'S LAW FOR DARK ENERGY}

\author{Paul H. Frampton\footnote{frampton@physics.unc.edu}}

\affiliation{Department of Physics and Astronomy, University of North Carolina,\\
Chapel Hill, NC 27599-3255, USA\footnote{Permanent address}\\
and\\
Centro Universitario Devoto, Buenos Aires, Argentina}

\date{\today}
\begin{abstract}
\begin{center} \textbf{Abstract} \end{center}A model is introduced in which Newton's law is modified between matter and dark energy corpuscles (DECs). The model predicts that the DEC component is
presently decelerating in its expansion at $14\%$ of the magnitude of the matter expansion acceleration. In the future, expansion of the DEC universe will continue to decelerate.
\end{abstract}

\maketitle

\newpage
The nature of dark energy \cite{1}-\cite{2} is one of the most pressing problems in
physics and astronomy. One underlying question is whether it necessitates a modification
of general relativity at large cosmological distances greater than $10^{12}$ m, the size of the solar system. If we are content with a cosmological constant term, as introduced by Einstein for an obsolete
reason, then general relativity survives. On the other hand, valiant attempts to make a microscopic theory of dark energy remain unconvincing.

Dark energy is a new driving term in the Friedmann equations doctored to fit the observed accelerated expansion. If the equation of state is $w = p/\rho = -1$ it coincides with a cosmological constant. Therefore
it is widely assumed that the future evolution will lead to an eternal exponential expansion with
acceleration. This assumption will be questioned in this article in which we will consider a model in which the repulsive nature of the associated force is accommodated more directly, and eventual deceleration
is inevitable.

The idea is to introduce a constituent particle of dark energy which we shall call a Dark Energy Corpuscle (DEC), and to postulate how it interacts with itself and with matter, by which we include both luminous and dark matter.

We recall that Friedmann's equations in a flat matter-dominated universe coincide with Newton's law of gravity. In the FLRW cosmology with vanishing curvature, and including both matter density $\rho_m$ and
dark energy $\rho_\Lambda$ they are
\begin{equation}
\left(\frac{\dot a}{a}\right) = \frac{8\pi G}{3} (\rho_m + \rho_\Lambda)
\label{1}
\end{equation}
and
\begin{equation}
\left(\frac{\ddot a}{a}\right) = -\frac{4\pi G}{3} (\rho_m + \rho_\Lambda + 3 P_\Lambda)
\label{2}
\end{equation}
where $a(t)$ is the scale factor and $P_\lambda$ is the dark energy pressure.

According to WMAP7 \cite{3}, the cosmology is consistent with zero curvature $k=0$ and critical energy
density divided into matter $\Omega_m = 0.28$ and dark energy $\Omega_\Lambda = 0.72$. The outward acceleration of a matter test particle is observed to be
\begin{equation}
a_m(t) = 13.40  \times 10^{-10} \,{\rm m/s^2}
\label{3}
\end{equation}
In order to break down Eq.(3) into the contributions of matter and of DECs,  we use for the critical density at the present time $t= t_0$
\begin{equation}
\rho_c(t_0) = 9.46 \times 10^{-27} \,{\rm kg \,m}^{-3}
\end{equation}
so that the matter and dark energy densities are
\begin{eqnarray}
\rho_m(t_0) &=& \Omega_m \rho_c(t_0) = 2.65 \times 10^{-27} \,{\rm kg \,m}^{-3}\label{5}\\
\rho_\Lambda(t_0) &=& \Omega_\Lambda\rho_c(t_0) = 6.81 \times 10^{-27} \,{\rm kg \,m}^{-3}\label{6}
\end{eqnarray}
From the same source \cite{3} the accurate comoving radius of the visible universe
bounded by the surface of last scatter (SLS) is
\begin{equation}
R_u(t_0) = 4.36 \times 10^{28} \,{\rm m}
\label{7}
\end{equation}
The corresponding masses of matter ($m_m(t_0)$) and dark energy ($m_\Lambda(t_0)$) from the densities Eqs.
(\ref{5}) and (\ref{6}) are
\begin{eqnarray}
{\cal M}_m(t_0) &=&   9.10 \times 10^{53} \,{\rm kg} = 4.55 \times 10^{23} M_\odot\label{8}\\
{\cal M}_\Lambda(t_0) &=&  2.37 \times 10^{54} \,{\rm kg} = 1.18 \times 10^{24} M_\odot\label{9}
\end{eqnarray}
using for the solar mass $M_\odot = 2 \times 2^{30}$ kg.

Let us consider a test particle of mass $(m)$ type at the SLS. Then, according to Newton's law its outward acceleration is, due to ${\cal M}_m(t_0)$,
\begin{equation}
\left( a_m(t_0)\right)_m = - \frac{G{\cal M}_m(t_0)}{R_u^2(t_0)} = - 3.22 \times 10^{-10} \,{\rm m/s^2}
\label{10}
\end{equation}
which was the expected prediction before the observation \cite{1,2} that $a_m(t_0)$ is given instead by
the very different value in Eq.(\ref{3}) having the opposite sign and much larger magnitude.

In the present DEC-model, we postulate a novel repulsive counterpart of Newton's law as the appropriate gravitational interaction between DECs and matter
\begin{equation}
F^{Newton}_{ m\!-\!DEC} = -\frac{2Gm_1m_2}{r^2}
\label{11}
\end{equation}
This provides an additional contribution to $a_m(t_0)$ given by
\begin{equation}
\left( a_m(t_0) \right)_\Lambda = \frac{G{\cal M}_\Lambda(t_0)}{R_u^2(t_0)} = +16.62 \times 10^{-10} \,{\rm m/s^2}
\label{12}
\end{equation}
The result for $a_m(t_0)$ is therefore outward by
\begin{eqnarray}
a_m(t_0) &=& \left(a_m(t_0)\right)_m + \left(a_m(t_0)\right)_\Lambda \label{13}\\
& = &
 +13.40 \times 10^{-10} \,{\rm m/s^2}
 \label{14}
\end{eqnarray}
which agrees exactly with observations in \cite{1,2} and Eq.(\ref{3}) above.

To complete our DEC model it is needed only to postulate the interaction between DECs and as the most conservative possibility we assume the same force as between matter particles, namely
\begin{equation}
F^{Newton}_{ DEC\!-\!DEC} = \frac{Gm_1m_2}{r^2}
\label{15}
\end{equation}
Since dark energy has never been directly observed, only indirectly from the expansion of the universe, and indirectly from other cosmological observations, it may seem challenging to confirm Eq.(\ref{15})

  The DEC model, however, provides a novel view of the future of the universe.

  From the Friedmann equation, Eq.(\ref{1}), and having in mind that the different scaling between $\rho_m \sim a^{-3}$ and $\rho_\Lambda \sim$\,constant the distant future evolution of the scale factor is
  \begin{equation}
 a(t) = \exp\left(H_0(t-t_0)
 \right)
 \label{16}
  \end{equation}
  where $(\dot a/a)$ is the Hubble parameter which is asymptotically constant.

  The behavior, Eq.(\ref{16}), suggests that the expansion of the visible universe continues to accelerate
  indefinitely.

  This conclusion is, however, misleading as follows. To explain perspicuously, it is first necessary to discuss the modest expansion of the visible universe from $R_u(t_0)$ to $R_u(t)$, $t \gg t_0$, starting from
  \begin{equation}
  R_u(t_0) = c \int_0^{t_0} \frac{dt}{a(t)} = 4.36 \times 10^{28} \,{\rm m }
  \label{17}
  \end{equation}
  to arrive at
  \begin{eqnarray}
  R_u(t_1) &=& 4.36 \times 10^{28} \,{\rm m } +  c \int_{t_0}^{t_1} \frac{dt}{a(t)}\label{18}\\
  &=&  4.36 \times 10^{28} \,{\rm m } + c H_0^{-1} \label{19}\\
  &=& 5.67 \times 10^{28} \,{\rm m }
  \label{20}
  \end{eqnarray}
  for any $t_1 \gg t_0$.

  As an illustration, let us choose the future redshift as
  \begin{equation}
  z_1 = -1 + 10^{-8}
  \label{21}
  \end{equation}
  so that $a(t_1) = 10^{-8}$ and the fraction of matter reduces to
  \begin{equation}
  \Omega_m(t_1) = 2.8 \times 10^{-25}
  \label{22}
  \end{equation}
  while the fraction of dark energy is $\Omega_\Lambda = (1 - \Omega_m)$.

  Because of the increase in the scale factor, no other galaxy will be visible \cite{4} so that normal
  matter will not be of interest to cosmologists. Instead, in the then visible universe it will be the DECs which
  will be the subject of scientific enquiry and detectable by the technology.

  For DECS the effect of the normal matter will be negligible and so a DEC test particle at the SLS will have an outward acceleration
  \begin{eqnarray}
  a_\Lambda(t_1) &=& \frac{G {\cal M}_\Lambda(t_1)}{R_u(t_1)^2}
  \label{23}\\
  &=&  10.77 \times 10^{-10} {\rm m/s^2}
  \label{24}
  \end{eqnarray}
  which will remain essentially constant in this DEC.

  Thus, the universe reverts to a decelerating expansion with respect to the overwhelming component of dark energy.

  Because the radius of the visible universe remains constant, all of the matter component exit to be rendered invisible \cite{4}. All that remains which is of cosmological interest are the dark energy corpuscles which by then we will be able to detect and measure. With regard to these, the cosmic expansion will be decelerating indefinitely into the infinite future even for the case of
  a constant dark energy equation of state $w_\Lambda = p_\lambda/\rho_\lambda = - 1$.

  It is mildly titillating to speculate that future technology may harness the gravitational repulsion
  in Eq.(\ref{11}) so that DECs fulfil the antigravity hopes mainly of science fiction writers, although
  antigravity does appear in some versions of supergravity theory \cite{5}.

  ~

\noindent{\bf Acknowledgments:}
I am grateful to J.~M.~Maldacena  for a useful discussion and to the Centro Universitario Devoto for providing the luxury of  unencumbered time. This work was supported by the U.S. Department of Energy Grant No. DE-FG02-05ER41418.

\end{document}